\documentclass[conference]{IEEEtran}
\IEEEoverridecommandlockouts

\usepackage{cite}
\usepackage{amsmath,amssymb,amsfonts}
\usepackage{algorithmic}
\usepackage{graphicx}
\usepackage{textcomp}
\usepackage{xcolor}
\usepackage{algorithm}
\usepackage{algorithmic}
\usepackage{booktabs}   
\usepackage{multirow}   

\usepackage{amsmath}
\usepackage{amssymb}
\usepackage[table,dvipsnames]{xcolor}
\definecolor{stepA}{RGB}{235,245,255}  
\definecolor{stepB}{RGB}{240,255,240}  
\definecolor{stepC}{RGB}{255,245,235}  
\usepackage{tikz}
\usetikzlibrary{
    arrows.meta,
    positioning,
    calc,
    shapes,
    fit,
    backgrounds
}
\usepackage{paralist}

\usepackage{float}
\def\BibTeX{{\rm B\kern-.05em{\sc i\kern-.025em b}\kern-.08em
    T\kern-.1667em\lower.7ex\hbox{E}\kern-.125emX}}
\begin{document}

\title{MOAT: \underline{Mo}del-\underline{A}gnostic Randomized \underline{T}ransformations to Prevent Efficiency Degradation Attacks on Vision Transformers\\}

\author{
\IEEEauthorblockN{
Anadi Goyal\IEEEauthorrefmark{1},
Nandish Chattopadhyay\IEEEauthorrefmark{1},
Chandan Karfa\IEEEauthorrefmark{1},
Anupam Chattopadhyay\IEEEauthorrefmark{2},
and Norrathep Rattanavipanon\IEEEauthorrefmark{3}
}

\IEEEauthorblockA{
\IEEEauthorrefmark{1}
Indian Institute of Technology Guwahati, India
}

\IEEEauthorblockA{
\IEEEauthorrefmark{2}
Nanyang Technological University, Singapore
}

\IEEEauthorblockA{
\IEEEauthorrefmark{3}
Prince of Songkla University, Phuket Campus
}
}
\maketitle

\begin{abstract}
To adopt the Vision Transformers (ViTs) in resource-constrained environment, token pruning \cite{ats,adavit} is widely used to reduce computational cost without impacting accuracy. 
However, adversaries have developed targeted attacks against said token pruning techniques to undermine such attempts to make ViTs efficient \cite{desparsify, slowformer}. In this paper, we propose \textbf{MOAT} (\textbf{MO}del \textbf{A}gnostic randomized \textbf{T}ransformations), a model-agnostic pre-processing defense pipeline that applies a combination of input transformations to protect efficient ViT implementations against adversarial efficiency attacks. MOAT operates directly on the input without requiring modifications to the model architecture or token pruning mechanism. Experimental results demonstrate that, across all evaluated ViT models, MOAT limits GFLOPs degradation under adversarial attacks to within 3.4\% of the original unattacked model.
\end{abstract}


\section{Introduction}
The Vision Transformer (ViT) \cite{50650} has rapidly gained widespread adoption in computer vision, emerging as a strong alternative to Convolutional Neural Networks (CNNs). Despite their effectiveness, ViTs suffer from high computational cost, as self-attention scales quadratically with the number of input tokens, limiting their deployment in resource-constrained environments. To
address this, researchers have developed a various optimization techniques for improving their efficiency. A prominent class of such techniques is input-adaptive token pruning \cite{ats,adavit}, which dynamically adjusts the number of retained tokens based on token importance for each input during inference. As illustrated in Figure~\ref{fig:placeholder} (green box), tokens assigned low importance are pruned (shown as white patches), resulting in reduced computational cost.
However, these efficiency-oriented \textit{Adaptive-ViTs} introduce new security vulnerabilities. Recent studies \cite{desparsify, slowformer} show that adversaries can exploit token pruning mechanisms present in Adaptive-ViTs to launch adversarial efficiency degradation attacks. Unlike traditional adversarial attacks that aim to alter model predictions, these attacks target the computational behavior of the model. Figure~\ref{fig:placeholder} (red box) illustrates an efficiency-degradation adversarial attack on Adaptive-ViTs. During inference, the adversary perturbs the input with carefully crafted, visually imperceptible noise that manipulates the victim model’s token-importance estimation. As a result, Adaptive-ViTs are forced to retain unnecessary tokens, leading to a substantial increase in inference-time computation, latency, and energy consumption. 
DeSparsify~\cite{desparsify} is one such attack which is a single-image attack that crafts input-specific perturbations, while Slowformer~\cite{slowformer} produces a universal adversarial patch applicable across a dataset; both share the objective of forcing ViTs to retain more tokens and incur higher computational cost.
Such behavior poses serious risks in real-time and resource-constrained deployments (e.g., surveillance systems or edge devices).

\begin{figure}
    \centering
    \includegraphics[width=1\linewidth]{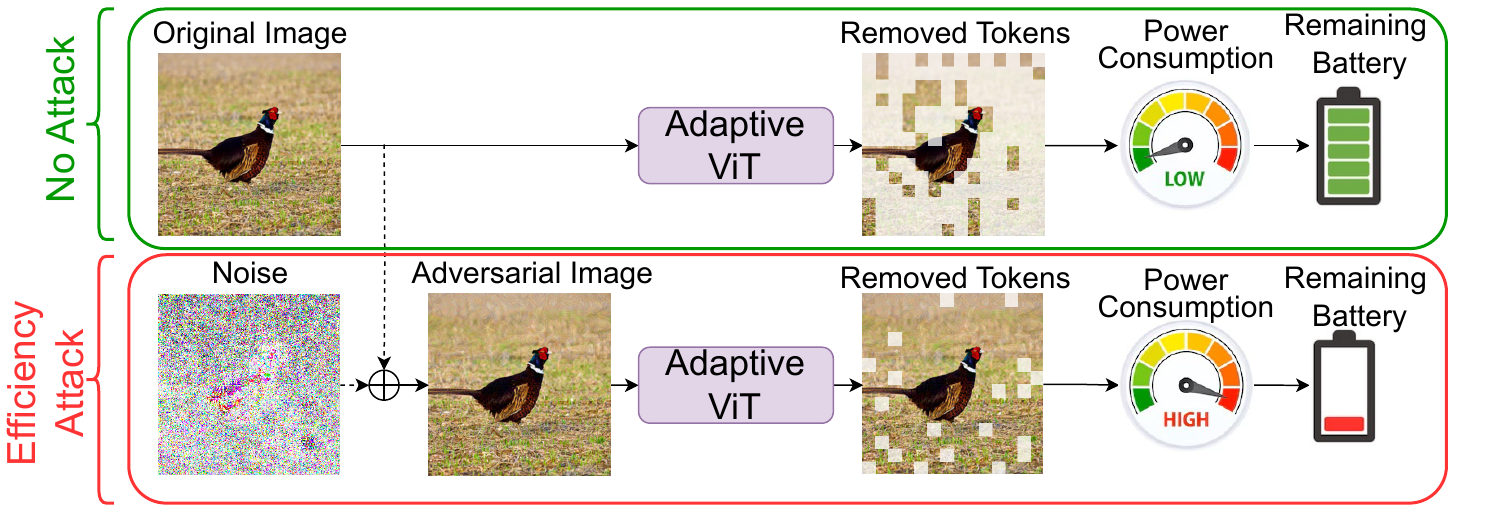}
    \caption{Efficiency-Degradation Attack on Adaptive Vision Transformers.}
    \label{fig:placeholder}
\end{figure}

{\it Related Works:} These attacks highlight the need for effective defenses against adversarial efficiency degradation in Adaptive-ViTs. While there are defenses against adversarial attacks \cite{cod_1, cod_2, cod_3, cod_4, cod_5}, to date, only two defense strategies have been explored in the domain of efficiancy degradation attacks: token capping \cite{desparsify} and adversarial training \cite{slowformer}.
Token capping enforces a fixed upper bound on the number of tokens retained during inference, ensuring that a minimum number of tokens are always removed. While this limits attack strength by preventing excessive token retention, it introduces a static constraint that reduces input adaptivity and may over-prune informative tokens in complex images.
Adversarial training improves robustness by retraining the model on adversarially crafted inputs. However, it provides only partial robustness, does not fully restore efficiency, and incurs substantial training cost in terms of time and computational resources.
Moreover, both defenses require modifications to the model or pruning mechanism, limiting practicality and deployment flexibility.

\begin{figure}[!htbp]
    \centering
    \includegraphics[width=1\linewidth]{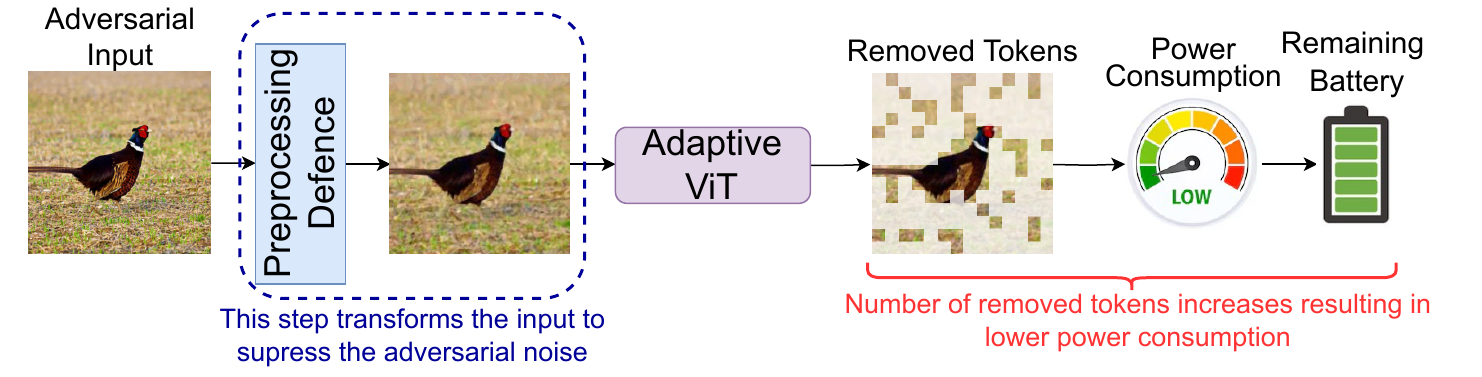}
    \caption{Overview of our Proposed Preprocessing Defence:\textbf{ MOAT}.}
    \label{fig:defence}
\end{figure}

\textbf{Contributions:} To address these limitations, this work proposes \textbf{MOAT} (\textbf{MO}del \textbf{A}gnostic randomized \textbf{T}ransformations), a preprocessing-based defense against efficiency degradation attacks as illustrated in  Figure~\ref{fig:defence}. 
MOAT operates directly on the input image using lightweight transformations, without modifying the model architecture or its token pruning strategy. These transformations aim to suppress adversarial noise, resulting in token importance patterns that closely resemble those of the clean input. Consequently, unnecessary token retention is avoided, and the inference efficiency degraded by the adversarial attack is largely restored.
The proposed defense consists of a sequential pipeline of three input transformations: random resizing, median filtering, and JPEG compression. While individual transformations provide limited protection, their combination is substantially more effective, as each targets different characteristics of adversarial noise. 
This layered design also makes it more difficult for an attacker to construct adaptive attacks that simultaneously bypass all transformations. Preliminary experiments indicate that our approach can recover a significant portion of the efficiency lost under the DeSparsify attack \cite{desparsify} on the ATS optimization method \cite{ats}, while introducing minimal overhead and limited impact on clean-time efficiency. To the best of our knowledge, \textit{MOAT is the \textit{first preprocessing-based defense against efficiency degradation attacks}.} 
\vspace{-0.6cm}
\section{Background \& Related Work}
\subsection{Token Pruning in ViTs}
Vision Transformers (ViTs) \cite{dosovitskiy2020image} operate by partitioning an image into patches and mapping them into embedding vectors, which are then processed through a sequence of encoder blocks comprising Multi-head Self-Attention (MSA), Multi-Layer Perceptrons (MLPs), and Layer Normalization (LN), before reaching the classification head. A patch denotes a spatial segment of the input image, whereas a token represents its corresponding embedding used during inference.
Token pruning refers to the process of selectively removing a subset of tokens during inference to reduce computational cost, while aiming to preserve model accuracy by retaining only the most informative tokens.
Token pruning can broadly be divided into static and dynamic (adaptive) strategies. Static approaches discard tokens in an input-independent manner, which risks removing semantically important information. Methods in this category often rely on attention scores, particularly between the CLS token and other tokens, to decide retention \cite{lee2024vit} or employ trained token selectors \cite{rao2021dynamicvit}.
In contrast, dynamic token selection \cite{ats,adavit} adapts to each input instance, retaining tokens deemed informative for that specific image. This input-aware mechanism generally achieves better preservation of critical information.

\subsection{Efficiency Vulnerability of Dynamic Inference}
\label{subsec:efficiency-attack-avit}

In this subsection, we explain how input-adaptive optimization methods designed to reduce computation can themselves be exploited to \emph{increase}, or even \emph{reverse}, the intended efficiency benefits.

Input-adaptive methods employ a \emph{compute policy} that maps an input image to a specific computation pattern. Let $\pi(x)$ denote this policy (e.g., token masks or halting decisions), and let $C(x)$ represent the corresponding computational cost in terms of FLOPs, latency, or energy. A model is considered \emph{efficiency-vulnerable} if there exists a perceptually small perturbation $\delta$, satisfying $\lVert \delta \rVert_p \le \varepsilon$, such that the perturbed input drives the computational cost toward a worst-case regime, i.e.,
\[
C(x+\delta) \approx C_{\text{dense}},
\]
while preserving the task output or causing only minimal degradation.

This vulnerability arises when the compute policy can be expressed as
\[
\pi(x)=\Phi\!\big(g(x)\big),
\]
where $g(x)$ consists of image-dependent differentiable signals such as attention maps, mask logits, halting scores, confidence margins, or expert-routing gates. By slightly perturbing these signals toward ``keep-more'', ``exit-later'', or ``activate-wider'' regions, an adversary can systematically increase the amount of computation performed during inference. Consequently, any input-adaptive optimization strategy that couples computation with smooth, content-dependent signals inherently exposes a similar vulnerability: the \emph{compute path} becomes manipulable from the input space in the same way that adversarial perturbations manipulate the \emph{decision boundary} in conventional adversarial attacks.

\subsection{Adversarial Efficiency Degradation Attacks: Past Works}
Only in recent years has research started investigating adversarial attacks targeting efficient neural networks. One of the earliest works, ILFO \cite{haque2020ilfo}, focused on two CNN-based input-adaptive architectures like SkipNet \cite{wang2018skipnet}, by generating image-dependent perturbations. However, ILFO exhibited instability in dynamic-width networks due to its sensitivity to gating decisions. To address this issue, GradAuto \cite{pan2022gradauto} introduced a multi-objective optimization framework that balanced gradient sensitivity across gates, producing more stable perturbations. It also proposed directional gradient optimization to minimize interference between active and inactive gates, thereby selectively increasing execution depth.
DeepSloth \cite{hong2020panda} explored attacks on early-exit networks by suppressing confidence scores at intermediate classifiers, preventing early exits and forcing full-depth inference. NICGSlowDown \cite{chen2022nicgslowdown} and TransSlowDown \cite{chen2022transslowdown} expanded adversarial efficiency degradation attacks to sequence-based applications, including neural image captioning and machine translation.
However, in this work, we concentrate on Adversarial Efficiency Degradation Attacks targeting ViTs \cite{desparsify}, with particular emphasis on the widely adopted optimization strategy of token pruning. The mechanism of these attacks is described in detail in the following subsection.

\subsection{Efficiency-Degradation Attack on ATS}
To generate adversarial examples that degrade the efficiency of Adaptive Token Sampling (ATS) \cite{ats}, DeSparsify attack \cite{desparsify} employ an $\ell_\infty$-bounded Projected Gradient Descent (PGD) optimization with a custom objective designed to decrease efficiency. At iteration $t$, the perturbation $\delta$ is updated as:

$\delta^{t+1} = \Pi_{\|\delta\|_\infty < \epsilon}
\Big(\delta^t + \alpha \cdot \mathrm{sign}(\nabla_\delta L(x,y)) \Big)$

where $\Pi$ denotes projection onto the $\ell_\infty$ ball of radius $\epsilon$, and where $L$ denotes the overall custom loss function designed for the efficiency-degradation attack objective.

The overall loss $L$ is composed of two terms:
$L = L_{\text{ATS}} + \lambda L_{\text{cls}}$,
where $L_{\text{ATS}}$ drives the efficiency-degradation objective and $L_{\text{cls}}$ preserves the original classification to ensure stealthiness.

\vspace{4pt}
\noindent\textbf{ATS Token Importance.}
In ATS, token pruning is governed by token significance scores derived from the attention of the classification token. For a token $j$, the score is computed as
\begin{equation}
S_j = 
\frac{A_{1,j} \cdot \|V_j\|}
{\sum_{i=2}^{N+1} A_{1,i} \cdot \|V_i\|},
\end{equation}
where $A_{1,j}$ denotes the attention weight from the CLS token to token $j$, and $V_j$ is its value vector. Since $S$ is normalized, it forms a probability distribution used for sampling tokens via the cumulative distribution function (CDF). Tokens with higher $S_j$ are sampled multiple times, resulting in fewer unique tokens retained after sampling.

\vspace{4pt}
\noindent\textbf{Attack Objective.}
To prevent token sparsification, the attack aims to maximize the number of unique sampled tokens. This occurs when the distribution $S$ is uniform, causing each token to be sampled exactly once. Therefore, the attack pushes $S$ toward a uniform distribution $\hat{S}$ using the KL-divergence:

\begin{equation}
L_{\text{ATS}} = \frac{1}{L} \sum_{l=1}^{L}
\mathrm{KL}\left(S^{(l)} \,\|\, \hat{S}\right),
\end{equation}

where $S^{(l)}$ is the token score distribution at transformer block $l$. Minimizing this loss forces balanced token importance across all tokens, resulting in maximal token retention and worst-case inference computation.

\begin{algorithm}[h]
\caption{Preprocessing-Based Defense Against Efficiency Degradation Attacks}
\label{alg:defense}
\begin{algorithmic}[1]
\REQUIRE Input image $x \in [0,1]^{3 \times H \times W}$
\ENSURE Defended image $\tilde{x}$

\STATE \colorbox{stepA}{%
\parbox{\linewidth}{%
\strut Sample random scale $s \sim \mathcal{U}(s_{\min}, s_{\max})$ \hfill (random spatial scaling)\\
\strut $x_1 \leftarrow \textbf{Resize}(x, s)$\\
\strut $x_1 \leftarrow \textbf{CenterCropOrPad}(x_1, H, W)$
}}

\STATE \colorbox{stepB}{\strut $x_2 \leftarrow \textbf{MedianFilter}(x_1, k)$ \hfill (kernel size $k$)}

\STATE \colorbox{stepC}{\strut $x_3 \leftarrow \textbf{JPEGCompress}(x_2, q)$ \hfill (quality $q$, DCT-based)}

\RETURN $\tilde{x}$
\end{algorithmic}
\end{algorithm}

\section{Proposed Preprocessing Defense}
\begin{figure}[t]
\centering
\begin{tikzpicture}[
    node distance=6mm and 0mm,
    every node/.style={font=\small},
    initblock/.style={
        draw=blue!60, fill=blue!8, rectangle,
        minimum width=6.6cm, minimum height=12mm, align=center,
        rounded corners=4pt, line width=0.8pt
    },
    vitblock/.style={
        draw=violet!70, fill=violet!8, rectangle,
        minimum width=6.6cm, minimum height=24mm, align=center,
        rounded corners=4pt, line width=0.8pt
    },
    lossblock/.style={
        draw=orange!70, fill=orange!7, rectangle,
        minimum width=7.4cm, minimum height=44mm, align=center,
        rounded corners=4pt, line width=0.8pt
    },
    gradblock/.style={
        draw=orange!80!black, fill=yellow!10, rectangle,
        minimum width=6.6cm, minimum height=14mm, align=center,
        rounded corners=4pt, line width=0.8pt
    },
    updateblock/.style={
        draw=teal!70, fill=teal!8, rectangle,
        minimum width=7.4cm, minimum height=14mm, align=center,
        rounded corners=4pt, line width=0.8pt
    },
    inputnode/.style={
        draw=gray!60, fill=gray!10, rectangle,
        minimum width=6.6cm, minimum height=11mm, align=center,
        rounded corners=4pt, line width=0.8pt
    },
    outputnode/.style={
        draw=red!50, fill=red!7, rectangle,
        minimum width=6.6cm, minimum height=11mm, align=center,
        rounded corners=4pt, line width=0.8pt
    },
    arrow/.style={->, thick, >=stealth},
    looparrow/.style={->, thick, >=stealth, rounded corners=10pt,
                      dashed, draw=gray!70}
]

\node[inputnode] (input)
    {\textcolor{gray!70!black}{\textbf{Clean image}} $x$};

\node[initblock, below=of input] (init)
    {\textbf{\textcolor{blue!60!black}{Initialize}}\quad
     $x^{\mathrm{adv}}_0 = x$};

\node[vitblock, below=of init] (vit) {
    \textbf{\textcolor{violet!80!black}{Efficient ViT inference} $f(x)$}\\[3pt]
    Patch $\!\rightarrow\!$ Tokens $\!\rightarrow\!$ Token selection
    $\!\rightarrow\!$ Pruning\\
    $\rightarrow$ Transformer $\rightarrow$ Output\\[2pt]
    \footnotesize\textit{(ATS-based adaptive token sampling)}
};

\node[lossblock, below=of vit] (loss) {
    \textbf{\textcolor{orange!80!black}{Attack objectives}}\\[5pt]
    \textit{Obj 1 — Maximise computational cost}\\
    $\displaystyle
      L_{\mathrm{ATS}} = \frac{1}{L}\sum_{l=1}^{L}
      \mathrm{KL}\!\left(S^{(l)}\,\|\,\hat{S}\right)$\\[6pt]
    \rule{6cm}{0.3pt}\\[4pt]
    \textit{Obj 2 — Preserve classification accuracy}\\
    $L_{\mathrm{cls}} = \mathrm{CE}\!\left(f(x{+}\delta),\,f(x)\right)$\\[6pt]
    \rule{6cm}{0.3pt}\\[4pt]
    \textbf{Final loss:}\quad
    $\mathcal{L} = L_{\mathrm{atk}} + \lambda\,L_{\mathrm{cls}}$
};

\node[gradblock, below=of loss] (grad)
    {\textbf{\textcolor{orange!70!black}{Compute gradient}}\quad
     $\nabla_x \mathcal{L}$};

\node[updateblock, below=of grad] (update) {
    \textbf{\textcolor{teal!70!black}{Perturbation update}}\\[2pt]
    $x^{\mathrm{adv}}_{t+1} =
     \Pi_{x,\varepsilon}\!\left(
       x^{\mathrm{adv}}_t + \alpha\,\mathrm{sign}(\nabla_x \mathcal{L})
     \right)$
};

\node[outputnode, below=8mm of update] (output)
    {\textbf{\textcolor{red!60!black}{Adversarial image}}
     $x^{\mathrm{adv}}$};

\draw[arrow, draw=blue!50]   (input.south)  -- (init.north);
\draw[arrow, draw=blue!50]   (init.south)   -- (vit.north);
\draw[arrow, draw=violet!60] (vit.south)    -- (loss.north);
\draw[arrow, draw=orange!60] (loss.south)   -- (grad.north);
\draw[arrow, draw=orange!60] (grad.south)   -- (update.north);
\draw[arrow, draw=teal!60]   (update.south) -- (output.north);

\draw[looparrow]
    (update.east) -- ++(18mm,0)
    |- node[pos=0.25, right, font=\footnotesize,
            text=gray!70!black] {Iterate}
    (vit.east);

\end{tikzpicture}
\caption{Overview of the adversarial efficiency-degradation attack.
The dashed loop feeds the updated perturbation back into inference
for the next iteration.}
\label{fig:desparsify_attack}
\end{figure}
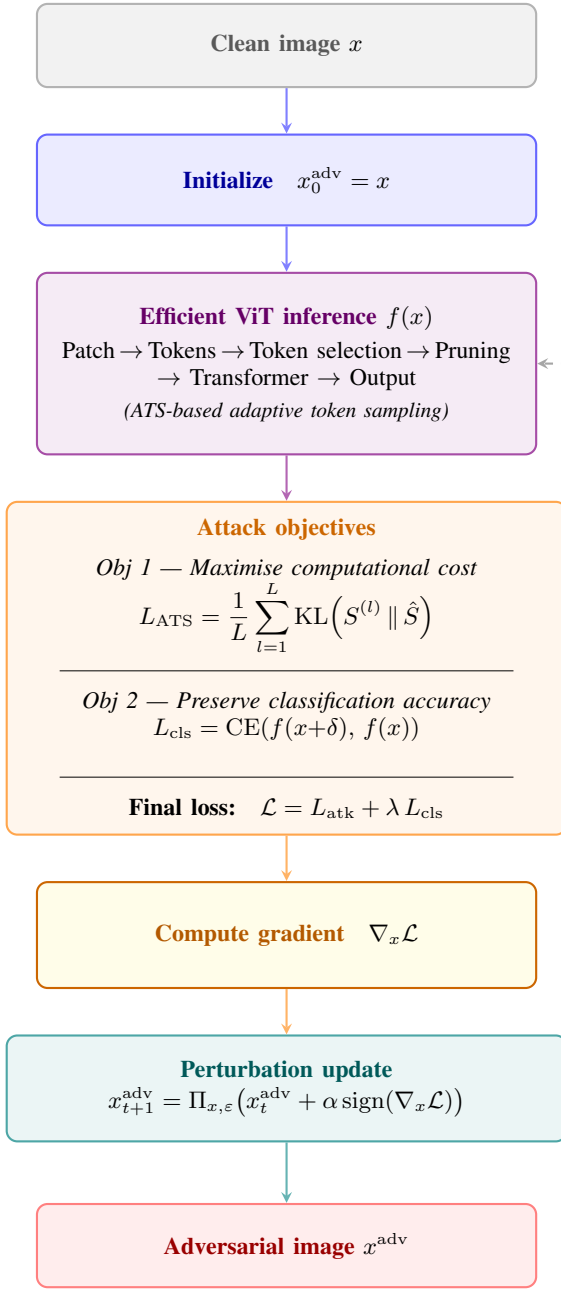
\subsection{Threat Model}
We consider a white-box adversary targeting input-adaptive token pruning mechanisms in Vision Transformers (ViTs). The complete attack pipeline is illustrated in Figure~\ref{fig:desparsify_attack}. The primary objective of the adversary is to generate an adversarial perturbation $\delta$ for an input image $x$ such that the token pruning mechanism retains the maximum number of tokens during inference, thereby increasing the computational overhead of the model. As a secondary objective, the attacker attempts to preserve the original model prediction in order to maintain the stealthiness of the attack.
To achieve these objectives, two loss functions, namely $L_{\text{ats}}$ and $L_{\text{cls}}$, are defined to respectively optimize token retention and classification consistency. During each optimization step, gradients of the combined loss are computed through a forward and backward pass of the model. The adversarial image is then iteratively updated using the optimization rule shown in Figure~\ref{fig:desparsify_attack}. This iterative process continues until the objective function converges.To evaluate the vulnerability of token pruning frameworks, we assume that the attacker has full knowledge of the victim model and its pruning strategy. We focus on a single-image attack setting, where a distinct perturbation $\delta$ is optimized for each input sample $x$.
\subsection{Overview of MOAT} 
Our defense MOAT is a preprocessing-based input transformation pipeline designed to mitigate efficiency degradation adversarial attacks targeting input-adaptive token pruning in ViTs. As shown in Algorithm \ref{alg:defense}, the defense operates directly on the input image and does not modify the model architecture or the pruning mechanism. The pipeline consists of three sequential transformations: random resizing, median filtering, and JPEG compression. Each transformation targets different characteristics of adversarial perturbations used in efficiency adversarial attacks, and their combination provides stronger robustness than any individual operation.
First, a random resizing is performed by sampling a scale factor $s \in [s_{\min}, s_{\max}]$, after which the image is center-cropped or padded back to its original size (line 1). This changes pixel coordinates and introduces randomness during inference.
Next, a median filter with kernel size $k$ is applied (line 2), where each pixel is replaced by the median value of its neighborhood, suppressing localized perturbations while preserving structural content.
Finally, JPEG compression with quality factor $q$ is applied (line 3). This DCT-based compression removes high-frequency components that often carry adversarial artifacts.
The resulting image $\tilde{x}$ is then passed to the ViT for inference.
\subsection{Rationale behind Each Transformation}

Efficiency degradation attacks increase token retention by injecting structured perturbations that manipulate token-importance estimation. The proposed transformations counter this behavior by introducing spatial randomness, suppressing localized high-frequency noise, and adding non-differentiability into the input processing pipeline.

\textbf{Random resizing} alters pixel coordinates and distorts the alignment between image pixels and patch embeddings. Since adversarial perturbations are crafted to influence attention scores at specific patch locations, even small changes in scale disrupt this precise alignment. Also, since the resizing factor changes randomly at inference time, it introduces additional uncertainty for the adversary. 

\textbf{Median filtering} removes localized, high-frequency perturbations by replacing each pixel with the median value of its neighborhood. When adversarial noise is concentrated within a small region (smaller than the kernel size), the median operation significantly reduces or eliminates its magnitude. At the same time, edges and larger structural patterns are preserved, ensuring minimal impact on clean images and meaningful tokens. 

\textbf{JPEG compression} removes residual high-frequency artifacts through DCT-based quantization while introducing non differentiability in the processing pipeline. This makes gradient-based crafting of adversarial noise more difficult.

Our proposed approach offers three key advantages over prior defenses \cite{desparsify, slowformer}:
(1) it is model-agnostic, requiring no modifications to the model architecture or token pruning mechanism,
(2) it is lightweight, introducing minimal inference overhead, which is critical when defending efficiency-oriented models, and
(3) it does not require any (re-)training, avoiding additional training cost.

\section{Results}
\subsection{Experimental Setup} 
\textbf{Models \& Datasets. }We evaluate our method on DeiT-Tiny and DeiT-Small, both pretrained on the ImageNet-1K dataset. All experiments are conducted on the ImageNet validation set, consisting of $500$ samples. 

\textbf{Attack Details.} We consider the \textit{DeSparsify} per-image noise--based efficiency-degradation attack \cite{desparsify} targeting the ATS token pruning framework \cite{ats}. The attack is implemented using PGD under an $\ell_\infty$ constraint of $\epsilon = 8/255$, with 500 iterations and a step size of $\epsilon/10$. The attack targets the ATS sparsification module applied to transformer blocks 4--12, with a classification-consistency regularizer ($\lambda = 8 \times 10^{-4}$). 

\textbf{Defense Details.} For defense, we employ a preprocessing pipeline consisting of random resize with a scale factor sampled from $[0.9, 1.1]$, median filtering with a $3 \times 3$ kernel, and JPEG compression with quality factor 75. All defense parameters are fixed across both models. Details regarding this choice of hyperparameters is discussed in later sections. 

\textbf{Metrics.} We report three evaluation metrics: the resulting GFLOPs, the \emph{classification accuracy}, and the \emph{Attack Success rate (ASR)}, which quantifies the extent of efficiency degradation. The ASR metric is computed as
\[
\mathrm{ASR} = \frac{F_{\text{attack}} - F_{\min}}{F_{\max} - F_{\min}} \times 100\%,
\]
where $F_{\min}$ denotes the GFLOPs of the corresponding framework under no attack, and $F_{\max}$ corresponds to the GFLOPs of the ViT-backbone model (with no token pruning).

\subsection{Evaluation} 
Table~\ref{tab:deit_results} summarizes the impact of DeSparsify adversarial attack and the proposed defense on accuracy (denoted as Acc.), GFLOPs, and ASR for DeiT-Small and DeiT-Tiny under different input conditions. Compared to clean inputs with ATS, adversarial examples (denoted as Adv.) substantially increase inference cost, reducing GFLOPs savings from 31.4\% to 12.4\% for DeiT-Small and from 34.8\% to 16.9\% for DeiT-Tiny.
Applying the proposed defense effectively restores efficiency under attack. For adversarial inputs, GFLOPs are reduced from 3.98 to 3.26 on DeiT-Small and from 1.05 to 0.87 on DeiT-Tiny, increasing savings to 28.2\% and 31.4\%, while lowering ASR to 7.5\% and 8.9\%. Importantly, the same defense configuration is applied to both models without modification, demonstrating the model-agnostic nature of the proposed approach. On clean inputs, the defense causes a drop in accuracy due to image-space transformations, but incurs negligible efficiency overhead, with GFLOPs remaining close to the clean ATS setting.
We further analyze the effectiveness of the combined defense pipeline relative to individual transformations. The full pipeline (Resize + Median + JPEG) consistently outperforms single-transformation defenses—where only one preprocessing operation is applied—by achieving a substantially lower attack success rate (7.5\%) compared to 15.4\%–32.1\% for individual defenses.

\begin{table}[!t]
\centering
\caption{Impact of efficiency-degradation attacks and defense on accuracy and GFLOPs. $^\dagger$Savings w.r.t.\ baseline GFLOPs.}
\label{tab:deit_results}
\small
\setlength{\tabcolsep}{3pt}
\begin{tabular}{llcccc}
\toprule
\textbf{Model} & \textbf{Scenario} & \textbf{Acc.} & \textbf{GFLOPs} & \textbf{Save$^\dagger$} & \textbf{ASR} \\
\midrule
\multirow{5}{*}{\rotatebox{90}{DeiT-Small}}
& Baseline              & 83.3\% & 4.54 & 0.0\%  & -      \\
& Clean+ATS             & 83.5\% & 3.12 & 31.4\% & -      \\
& \cellcolor{red!12}   Adv.+ATS
& \cellcolor{red!12}   65.1\%
& \cellcolor{red!12}   3.98
& \cellcolor{red!12}   12.4\%
& \cellcolor{red!12}   60.6\% \\
& \cellcolor{green!12} Adv.+Def.
& \cellcolor{green!12} 73.7\%
& \cellcolor{green!12} 3.26
& \cellcolor{green!12} 28.2\%
& \cellcolor{green!12} 7.5\% ($\downarrow$53.1\%) \\
& Clean+Def.            & 76.5\% & 3.16 & 30.5\% & -      \\
\midrule
\multirow{5}{*}{\rotatebox{90}{DeiT-Tiny}}
& Baseline              & 78.8\% & 1.23 & 0.0\%  & -      \\
& Clean+ATS             & 79.0\% & 0.83 & 34.8\% & -      \\
& \cellcolor{red!12}   Adv.+ATS
& \cellcolor{red!12}   55.4\%
& \cellcolor{red!12}   1.05
& \cellcolor{red!12}   16.9\%
& \cellcolor{red!12}   56.5\% \\
& \cellcolor{green!12} Adv.+Def.
& \cellcolor{green!12} 64.8\%
& \cellcolor{green!12} 0.87
& \cellcolor{green!12} 31.4\%
& \cellcolor{green!12} 8.9\% ($\downarrow$47.6\%) \\
& Clean+Def.            & 67.4\% & 0.83 & 34.1\% & -      \\
\bottomrule
\end{tabular}
\end{table}

\begin{table}[t]
\centering
\caption{Comparison of defence configurations across robustness and accuracy metrics.}
\label{tab:defence_comparison}
\setlength{\tabcolsep}{6pt}
\renewcommand{\arraystretch}{1.25}
{\scriptsize
\begin{tabular}{l c c c c}
\toprule
\multirow{2}{*}{\textbf{Type}}
    & \textbf{Hyperparameters}
    & \textbf{Clean}
    & \textbf{Adversarial}
    & \multirow{2}{*}{\textbf{ASR}} \\
    & \textbf{(range,\ $k$,\ $q$)}
    & \textbf{Accuracy}
    & \textbf{Accuracy}
    & \\
\midrule
More Robustness
    & {[0.70,\,1.30],\ 7,\ 30}
    & 51.4\%  & 50.0\%  & 2.3\% \\
More Accuracy
    & {[0.95,\,1.05],\ 3,\ 90}
    & 79.1\%  & 76.5\%  & 12.1\% \\
\textbf{Balanced}
    & \textbf{[0.90,\,1.10],\ 3,\ 75}
    & \textbf{76.5\%}
    & \textbf{72.6\%}
    & \textbf{7.5\%} \\
\bottomrule
\end{tabular}
}
\end{table}

\begin{figure}
    \centering
    \includegraphics[width=1\linewidth]{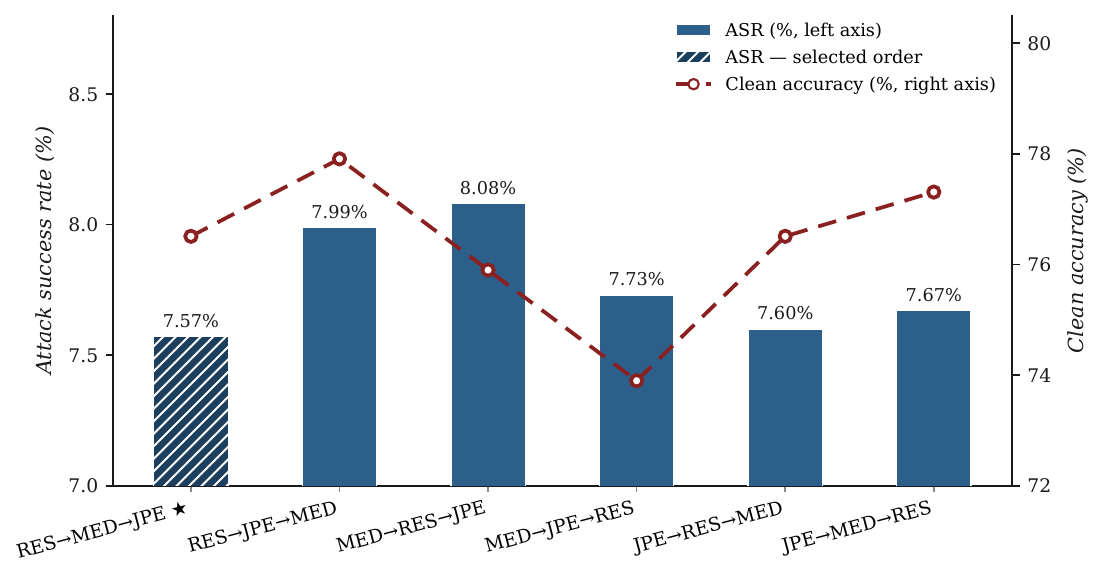}
    \caption{ASR and clean accuracy across all six defence orderings. Hatched bar indicates the selected configuration.}
    \label{fig:placeholder}
\end{figure}

\subsection{Choosing best hyperparameter}
Our defense consists of three transformation algorithms. The effectiveness of the framework depends on two factors: (1) the choice of hyperparameters for each transformation and (2) the ordering of the transformations. To balance robustness and clean-sample accuracy, we evaluated multiple hyperparameter settings and all possible transformation orders.

For random resizing, the primary hyperparameter is the resize-factor range $(s_{\min}, s_{\max})$; for median filtering, the kernel size $k$; and for JPEG compression, the quality factor $q$. We evaluated $s_{\min}\in[0.70,0.95]$ and $s_{\max}\in[1.05,1.30]$ with step size $0.05$, kernel sizes from $k=3$ to $11$, and JPEG quality factors from $20$ to $90$. Our experiments revealed the expected trade-off between robustness and clean accuracy: aggressive settings achieved very low ASR but reduced clean accuracy, while conservative settings preserved accuracy with weaker robustness.

Based on these observations, we selected the three representative configurations shown in Table~\ref{tab:defence_comparison}. The aggressive configuration achieves the lowest ASR ($2.3\%$) at the cost of reduced clean accuracy, while the accuracy-oriented configuration preserves clean performance with slightly weaker robustness. In this work, we adopt the balanced configuration, which provides a favorable trade-off between robustness and accuracy.

We additionally evaluated all six possible transformation orders, shown in Figure~\ref{fig:placeholder}. Although ordering had only a minor effect on performance, the sequence Random Resize $\rightarrow$ Median Filter $\rightarrow$ JPEG Compression (RES-MED-JPE) achieved the lowest ASR ($7.57\%$) while maintaining competitive clean accuracy, and was therefore selected for all experiments.

\begin{table}[h]
\centering
\caption{Comparison of inference cost: original model vs.\ DDPM Rectifier defense vs.\ our proposed defense. Overhead is expressed as a percentage of the original model GFLOPs.}
\label{tab:gflops_comparison}
\small
\setlength{\tabcolsep}{4pt}
\begin{tabular}{lcc}
\toprule
\textbf{Model} 
    & \textbf{+ DDPM Rectifier} 
    & \textbf{+ MOAT} \\
\textbf{(GFLOPs)}
    & \textbf{(GFLOPs / overhead)}
    & \textbf{(GFLOPs / overhead)} \\
\midrule
DeiT-Tiny \ (1.23)  & 595.23 \ / \ 48,270\% & 1.237 \ / \ 0.57\% \\
DeiT-Small (4.54)   & 598.54 \ / \ 13,079\% & 4.547 \ / \ 0.15\% \\
\bottomrule
\end{tabular}
\end{table}
\subsection{Overhead Analysis}
These attacks aim to target the efficiency of the models. Therefore, the defence should be chosen 
in a manner that not only retains efficiency by suppressing adversarial perturbations but also 
incurs low overhead and does not compound inefficiency. Our defence employs low-overhead 
transformations to satisfy both criteria. To highlight this, we compare the computational cost and overhead of a diffusion-based defense (DDPM) \cite{nie2022diffpure} with those of our lightweight defense. As shown in Table~\ref{tab:gflops_comparison}, while 
the DDPM-based Rectifier introduces a fixed cost of 594 GFLOPs regardless of 
the target model, resulting in overheads of 482$\times$ and 131$\times$ for DeiT-Tiny and 
DeiT-Small, respectively, MOAT adds only 0.007 GFLOPs. This corresponds to merely 
0.57\% and 0.15\% of the total computation for DeiT-Tiny and DeiT-Small, respectively, making 
it well-suited for efficiency-oriented ViTs.
\subsection{Discussion on Adaptive Attacks}

In a fully adaptive threat setting, the attacker is assumed to have complete knowledge of the defense pipeline, including all transformations and parameter settings. Our objective is to make adaptive attacks difficult by increasing their optimization difficulty and computational cost through two key design choices: \textit{randomness} and \textit{non-differentiability}.

\textbf{Randomness.}
Our defense introduces controlled randomness through random resizing, where the resize factor is randomly sampled during inference. This randomness destabilizes gradient estimation across forward passes, reducing the effectiveness of standard gradient-based attacks. An adaptive attacker would therefore require techniques such as \textit{Expectation over Transformation} (EOT), which estimates gradients over multiple stochastic evaluations, significantly increasing attack cost and generation time.
Although additional randomness could further increase attack difficulty, excessive stochasticity can negatively impact clean-sample accuracy. Prior studies have also reported that aggressive randomization often introduces noticeable degradation in model performance on benign inputs~\cite{random}. Hence, our framework employs moderate randomness to balance robustness and clean accuracy.

\textbf{Non-Differentiability.}
The pipeline also incorporates non-differentiable transformations that obstruct reliable gradient propagation. For example, JPEG compression applies quantization:
\[
g_{\text{JPEG}}(x) = \text{round}\left(\frac{x}{\Delta}\right)\cdot\Delta
\]
where $\Delta$ is the quantization step size. Since the rounding operation is non-differentiable, gradient-based optimization becomes less reliable, thereby increasing the difficulty of generating adversarial perturbations.

\vspace{-1mm}
\section{Future Work}
This work opens up a necessary direction of research aligned with protecting efficient implementations of ViTs like the token pruning approach, from adversaries who are equipped with tools to undermine said efficiency. This defense mechanism, with required modifications, need to be studied for physical and more potent patch based adversarial attacks across other machine learning objectives like object recognition and segmentation etc. Additionally, refinements of the algorithm can be designed for further reduction of loss in accuracy. 

\section*{Acknowledgment}
This paper is partially supported by ASEAN-India Collaborative Research Project entitled "A Toolchain for Secure Hardware Accelerator Design" (CRD/2024/000829).
\bibliographystyle{IEEEtran}
{\scriptsize\bibliography{sample-base}}

\end{document}